\documentclass[letterpaper,twocolumn,10pt]{article}
\usepackage{usenix-2020-09}

\usepackage{amsmath}
\usepackage{amssymb}
\usepackage{booktabs}
\usepackage{cite}
\usepackage{colortbl}
\usepackage{graphicx}
\usepackage{float}
\usepackage{tabularx}
\usepackage{textcomp}
\usepackage{tikz}
\usepackage{xcolor}

\begin{document}

\date{}

\title{\Large \bf Understanding How to Inform Blind and Low-Vision Users about Data Privacy through Privacy Question Answering Assistants}

\author{
{\rm Yuanyuan Feng}\\
University of Vermont
\and
{\rm Abhilasha Ravichander}\\
Allen Institute for AI
\and
{\rm Yaxing Yao}\\
Virginia Tech
\and
{\rm Shikun Zhang}\\
Carnegie Mellon University
\and
{\rm Rex Chen}\\
Carnegie Mellon University
\and
{\rm Shomir Wilson}\\
Pennsylvania State University
\and
{\rm Norman Sadeh}\\
Carnegie Mellon University
} 

\maketitle

\begin{abstract}
Understanding and managing data privacy in the digital world can be challenging for sighted users, let alone blind and low-vision (BLV) users.
There is limited research on how BLV users, who have special accessibility needs, navigate data privacy, and how potential privacy tools could assist them.
We conducted an in-depth qualitative study with 21 US BLV participants to understand their data privacy risk perception and mitigation, as well as their information behaviors related to data privacy.
We also explored BLV users' attitudes towards potential privacy question answering (Q\&A) assistants that enable them to better navigate data privacy information.
We found that BLV users face heightened security and privacy risks, but their risk mitigation is often insufficient. 
They do not necessarily seek data privacy information but clearly recognize the benefits of a potential privacy Q\&A assistant. 
They also expect privacy Q\&A assistants to possess cross-platform compatibility, support multi-modality, and demonstrate robust functionality.
Our study sheds light on BLV users' expectations when it comes to usability, accessibility, trust and equity issues regarding digital data privacy.
\end{abstract}

\section{Introduction} \label{sec:intro}
Navigating information about how websites, mobile applications, digital services, and Internet-connected devices (``digital technologies'' thereafter) collect, use, and share personal data is challenging.
First, it is difficult to find privacy policies -- the legal documents mandated by many privacy regulations that disclose data privacy practices~\cite{Sundareswara2020}.
Then, understanding the data practices disclosed in these privacy policies can be even more challenging due to their length, legal jargon, and vague language\cite{Mcdonald2008,Krumay2020}.
Given how difficult it is in general for people to find and understand data privacy information, it is natural to wonder how accessible these tasks are for people who are blind or have low vision.
According to the Cornell University Employment and Disability Institute's interpretation of the 2016 American Community Survey, more than seven million US adults (2.4\%) have a visual disability~\cite{BlindnessStats}.
This population often relies on assistive technology, such as screen readers and magnifiers, to access digital information.
In this paper, we use positive affirming adjectives recommended by the US National Federation of the Blind~\footnote{The National Federation of the Blind. \url{https://nfb.org/}}
---``blind'' and ``low-vision'' (instead of ``visually-impaired'')---to describe this population (``BLV people/users'' thereafter).

Prior research shows that BLV people are particularly vulnerable to online security and security threats due to the lack of visual cues~\cite{Ahmed2015, Hayes2019}. Generally, digital technologies tend to have poor accessibility features to support their general information needs in the digital world~\cite{Dosono2015,Kim2016}.
Therefore, it is critical to improve not only the usability but also the accessibility of security and privacy (S\&P) tools.
We seek to understand BLV people's needs in navigating the data privacy information regarding the digital technologies with which they interact.
We also aim to inform the design of accessible privacy tools, enabling BLV users to have the same level of access to privacy information as sighted users~\cite{zhao2023if}.

Specifically, we want to explore BLV users' attitudes towards ``privacy assistants'', broadly defined in this paper as tools designed to help users navigate and/or manage digital data privacy~\cite{liu2016follow,das2018personalized,Colnago2020}. Recently, there has been considerable research on developing privacy assistants that can answer users' privacy questions based on the content of privacy policies using natural language processing (NLP) techniques~\cite{harkous2018polisis, ravichander-etal-2019-question,Ravichander2021a}. Inspired by these recent advances in privacy question answering and the increasing public interest in applications such as ChatGPT\footnote{ChatGPT. \url{https://chat.openai.com/}} built on Large Language Models, we particularly investigate the NLP-based privacy question answering assistants (``privacy Q\&A assistants'' thereafter) as a promising approach to improve accessibility and bridge potential privacy inequity for BLV users. 

To this end, we conducted a qualitative interview study with 21 US BLV users of various digital technologies to investigate three research questions (RQs):
\begin{itemize}
\item \emph{RQ1: How do BLV people perceive and mitigate data privacy risks associated with digital technologies?} 
\item \emph{RQ2: What are BLV people's information (seeking) behaviors around data privacy?}
\item \emph{RQ3: What do BLV people expect from potential privacy Q\&A tools for navigating data privacy information?}
\end{itemize}

By answering these RQs, this paper contributes:
\begin{itemize}
    \item The first in-depth qualitative investigation into how BLV users perceive and mitigate data privacy risks and how they seek (if any) data privacy information.
    \item To the understanding of BLV users' expectations around functionality and accessibility for potential privacy Q\&A assistants that inform them about data privacy.
    \item To the growing area of inclusive security and privacy (S\&P), providing insights into designing accessible, usable, and equitable S\&P tools for BLV users and beyond.
\end{itemize}

\section{Background and Related Work}

\subsection{How BLV Users Access Information}
BLV users primarily rely on screen readers (e.g., JAWS and NVDA software on computers, VoiceOver on Apple devices, Talkback on Android devices), a type of assistive technology that reads out aloud text on the device screen, to access digital information and utilize other digital technologies.
Screen readers only function well when digital technologies and contents follow good accessibility practices.
However, research~\cite{Hackett2003,Acosta2019,Acosta2021} revealed that websites and mobile applications do not consistently adhere to accessibility standards such as the Web Content Accessibility Guidelines~\cite{Henry2005}. 

Recently, agent-based visual interpreter services, including Aira\footnote{Aira. \url{https://aira.io}} and Be My Eyes\footnote{Be My Eyes. \url{https://www.bemyeyes.com}} mobile apps, have gained traction among BLV users.
These services connect BLV users with sighted human agents, who help them recognize objects and cope with everyday situations through the phone cameras.
Similarly, artificial intelligence(AI)-based visual-aid apps like Seeing AI\footnote{Seeing AI. \url{https://www.microsoft.com/en-us/ai/seeing-ai/}} that leverage device cameras to describe texts, objects, people, and environments are also on the rise, enabling BLV users to access certain information independently.

Meanwhile, computing researchers have explored novel technical solutions and interaction modalities to improve information access for BLV people.
As smart home speakers and their built-in voice-based assistants (e.g., Amazon Echo devices with Alexa) gain popularity, voice user interfaces significantly improve information accessibility for BLV people~\cite{Pradhan2018}.
Also, research advances in computer vision~\cite{Gurari2018} and mixed reality~\cite{Zhao2016,Zhao2019} have also opened new possibilities to convey visual information through other modalities.
We broadly categorize these above-mentioned examples as assistive technology in this paper.

\subsection{Privacy and Security for BLV Users}
With growing recognition of the importance of accessibility and inclusiveness in privacy and security technologies~\cite{Wang2018}, many research studies examined blind and low-vision users' privacy concerns and behaviors.
Ahmed et al.~\cite{Ahmed2015}'s interview study revealed blind participants' unique privacy concerns in three environments: physical (e.g., eavesdropping), digital (e.g., privacy settings in social media), and the intersection of physical and digital (e.g., shoulder-surfing). 
To fulfill their privacy and security needs, Hayes et al. found that BLV users also rely heavily on their allies (e.g., family members, caregivers) to protect their privacy and security cooperatively~\cite{Hayes2019}. 
Akter et al.’s survey study elaborated on BLV users' concerns regarding camera-based visual interpreter services, where volunteers answer their questions about photos or videos~\cite{Akter2020}. 

Though focusing on a different population, Hamidi et al.'s study revealed that the privacy and utility trade-offs of adaptive assistive technologies might be overlooked among older adults with pointing problems~\cite{Hamidi2018}. This suggests that users with accessibility needs may knowingly use assistive technologies or services that compromise their data privacy. Such divergence between privacy attitudes and behaviors is commonly known as the privacy paradox~\cite{Gerber2018}.

Another important body of work focuses on the usability and accessibility of security and privacy tools for BLV users. For example, Danoso et al. found that web authentication can be time-consuming to BLV users and pose significant challenges, such as accessing error messages~\cite{Dosono2015,Dosono2018}. 
More importantly, these usability issues may result in BLV users' risky behaviors (e.g., not being able to identify phishing websites) or decisions that compromise security~\cite{Napoli2021}. 
Such usability issues also impact how BLV users seek and access general information~\cite{Bates2017,Tomlinson2016,Mutula2016,Xie2021}.
A study investigating online information behaviors revealed the barriers for blind web users to assess the credibility of websites~\cite{Abdolrahmani2016}, highlighting the need to assist BLV users in verifying the credibility of online information. 
However, to our knowledge, there is no research focusing on BLV users' information behaviors for data privacy information, a type of information that is challenging even for sighted users to navigate and understand~\cite{Sundareswara2020,Reidenberg2015,Vail2008,Oeldorf2019}.
Our study aims to explore this untrodden topic to understand the challenges faced by BLV users in navigating data privacy.

\subsection{Privacy Question Answering (Q\&A)}
Data privacy information about digital technologies can be obtained through many channels from different sources. A primary source is the official privacy policies -- the legal documents in which companies or organizations self-disclose the practices they engage in with user data.
Despite being required in many regulatory regimes around the world, privacy policies are difficult to find~\cite{Sundareswara2020} and even more difficult to understand due to their technical and legal jargon~\cite{Meiselwitz2013}. The time required to read them is also impractical for most users~\cite{Obar2020}.

Thus a growing line of research has explored using natural language processing (NLP) techniques to extract salient information from privacy policies, which can empower users to take control of their data privacy in the digital world. Prior studies~\cite{costante2012websites, ammar2012automatic, 10.1145/2381966.2381979, liu-etal-2014-step, ramanath2014unsupervised, wilson2016creation} have successfully identified data practices within privacy policies to potentially enable easier navigation of privacy policy content. Kumar et al.~\cite{Kumar2020} extracted opt-out choices from website privacy policies and presented them to users in a web browser extension, \textit{Opt-Out Easy}. 
Prior research efforts \cite{10.1145/2857705.2857741,zaeem2018privacycheck, zaeem2020privacycheck,keymanesh2020toward} also developed techniques to summarize the most salient aspects of privacy policies to present to end-users. However, other research has indicated that summarization approaches that are not tailored to the needs of individual users are unlikely to be effective at meaningfully informing people about the information in privacy policies \cite{gluck2016short, rao2016expecting}. Consequently, in recent years, there has been considerable interest in developing assistants that will answer users' privacy questions, which would enable people to flexibly access information within privacy policies that are most pertinent to them. 
Harkous et al.~\cite{Harkous2016} created \textit{PriBots}, chatbots for communicating privacy practices to users based on questions asked by users on Twitter. 
Ravichander et al.\cite{ravichander-etal-2019-question} constructed a benchmark for privacy question answering systems, where they source answers from experts with legal training and provide systems based on pretrained language models to identify answers to these questions. 
Ahmad et al. \cite{ahmad-etal-2020-policyqa} extracted segments from policies in the OPP-115 Corpus~\cite{wilson2016creation}, recruited ``skilled annotators'' to construct questions based on these segments, and explored transfer learning-based approaches to provide answers to these questions.

Building on existing research efforts to automatically answer users' data privacy questions~\cite{Ravichander2019,Ravichander2021b}, our study explores BLV users' expectations for similar privacy Q\&A tools to assist them in navigating data privacy information.
\section{Methods} \label{sec:method}

\subsection{Assumption and Method Justification} \label{sec:method:assumption}
Informed by literature review and guidance from two blind consultants, this study assumes that BLV people face challenges in navigating data privacy information and they can benefit from accessible privacy tools like the privacy Q\&A assistant.
To account for the potential inaccuracy of this assumption, we carefully formulated our research questions (RQs) in an open-ended manner, as shown in Section~\ref{sec:intro}).

We chose the qualitative interviewing method for its strengths in obtaining a deep understanding of participants' perceptions, attitudes, and experiences~\cite{Breakwell2006}, which is well-suited for RQ1 and RQ2.
Alternative methods to address RQ3 are prototype testing and evaluation with users, which are more beneficial during the late stages of software development~\cite{salovaara2017evaluation}.
Because there was little understanding of BLV users’ needs around privacy Q\&A, we decided not to present researcher-derived prototypes to avoid inaccurate assumptions about BLV users' preferences.
Instead, we decided to ask participants to freely imagine a hypothetical ``digital assistant'', a technique used in prior privacy and security interview studies~\cite{Colnago2020, Stover2023,yao2017folk} to elicit their requirements and expectations. This technique is particularly advantageous for requirements gathering in the early stages of tool design~\cite{Holtzblatt1995}.

\subsection{Study Design and Research Ethics} \label{sec:method:design}
We used the university-licensed Zoom software to conduct remote interviews in 2021 during the COVID pandemic.
We sought advice from two blind consultants in our academic network to ensure all study procedures are culturally responsible.
The first consultant provided us with best practices for recruiting BLV participants into research studies. The second consultant completed a mock-up interview and provided suggestions to improve the study materials. 
For example, we revised our study materials to accommodate varying technology literacy and the high unemployment rate among BLV people in the US. 
Then, we conducted a pilot interview with a blind friend to refine the wording of some interview questions and finalize all study materials.

Our study protocol was approved by the Institution Review Board (IRB) at Carnegie Mellon University with the permission to obtain informed consent verbally because our participants may have difficulty signing electronic documents. 
We emailed participants the IRB-approved consent form when we scheduled interviews because many BLV users prefer reading text via screen readers at a fast speed. 
When they joined the interview remotely, we asked if they had read the consent form. If yes, we read the study summary section of the consent form; if not, one researcher read the full consent form. Then we obtained their verbal consent before starting the audio recording. We also completed a research data sharing agreement after the first author joined the University of Vermont.

\subsection{Interview Questions} \label{sec:method:questions}
We started each interview by asking the participant's preferred terminology to describe their vision status so that we could use their preferred terminology throughout the interview.
We structured our interview questions based on the RQs, with additional baseline questions at the start and optional demographic questions at the end. We describe interview question design rationale below and provide them in Appendix ~\ref{sec:appendix:questions}.

\noindent \textbf{Baseline questions}: We started with questions to establish a baseline of participants' technology use, understanding of data practices, and attitude towards data privacy.
During baseline questions, we provided them with plain-language definitions for ``digital technologies'', ``data privacy'', and ``data practices'' in the context of this study to ensure a shared understanding of these terminologies (see Appendix~\ref{sec:appendix:def}). 
After providing the definitions, we asked about their general thoughts around data privacy because people's privacy attitudes are shown to impact their perceived risks~\cite{Kokolakis2017}.

\noindent \textbf{RQ1: Risk perception and mitigation.}
We asked participants to describe the potential risks associated with the digital technologies that they use and optionally to compare the risk levels between general and assistive technology.
We employed the critical incident technique~\cite{Flanagan1954} by asking participants to share their most memorable experiences (i.e., critical incidents) around data privacy in the past few months.
Then, we asked a hypothetical dilemma question to understand how they would mitigate privacy risks if assistive technology engaged in data practices that they were uncomfortable with.

\noindent \textbf{RQ2: Information (seeking) behaviors.}
Not assuming all participants actively seek data privacy information, we used ``information behaviors''~\cite{Bates2017} to broadly describe how participants navigate, access, seek, and understand data privacy information. 
We first asked about the sources from which they obtain data privacy information and their perceived credibility of these sources.
Then, we focused on their prior experience seeking data privacy information, if any.
We noted if participants naturally mentioned privacy policies in their responses. If not, we prompted privacy policies as a source.

\noindent \textbf{RQ3: Expectations for privacy Q\&A tools.}
To minimize the bias towards privacy Q\&A assistants as the researcher-introduced solution, we crafted the interview questions to emphasize participants' needs for privacy Q\&A.
We first introduced an imaginary privacy expert who can answer their data privacy questions for any digital technologies they use.
Then, we let them assume this expert was a ``digital assistant'' and asked them to freely describe their expectations for this digital assistant without priming them with any information about how the assistant might function.
After that, we probed further into detailed aspects of this hypothetical digital assistant, including modalities, developers, and information sources.
Finally, we asked about their perceived benefits, concerns, and potential use cases.

\noindent \textbf{Demographic questions:}
The interview ended with optional demographic questions.
All participants answered these questions to provide us with data on sample diversity.

\subsection{Recruitment and Data Collection} \label{sec:method:recruitment}
Recruiting BLV people into research studies requires trust-building, so we recruited participants through our personal contacts and the National Federation of the Blind.
Our consultants cautioned us about the sampling bias towards BLV users who are already comfortable using digital technologies. To mitigate this, we included a phone number in the recruitment materials and provided accessible instructions for joining Zoom via the Internet or phone call.

Aiming to increase sample diversity, we included the following statement in all study recruitment materials: ``We particularly welcome diverse perspectives from individuals who are less familiar with technology and who also belong to other underrepresented groups.'' 
When replying to study inquiries, we politely explain the rationale prioritizing participants from underrepresented groups. Many voluntarily shared their demographic information with us during inquiry, and those who were not chosen expressed their understanding.
Besides the above-mentioned diversity consideration, we responded to potential participants primarily by their time of inquiry. We aimed for a sample size of around 20 by referencing prior qualitative studies with BLV participants~\cite{Dosono2015,Ahmed2015,Pradhan2018,Hayes2019}. 

We recruited 21 self-identified blind or low-vision participants in the US: 2 from our personal connections with snowball sampling and 19 through the approved email solicitation of the National Federation of the Blind.
To avoid the potential discomfort some people feel when realizing they are the first few participants in a study, we assigned the number P10 to the pilot participant and numbered the full study participants from P11 to P31.
Note that we did not use data saturation~\cite{fusch2015we} to determine the sample size due to the inherent difficulty recruiting BLV users. Instead, we set our sample size goal by referencing published qualitative studies with this population.

The first and second authors conducted the first four interviews together and slightly adjusted interview question wording through discussing the initial results with the research team. All remaining interviews were led by either the first or second author, with an optional secondary interviewer from the research team.
We audio-recorded all interviews using Zoom after requesting participants to turn off their cameras during the interviews to ensure only audio was recorded.
We compensated each participant with an accessible electronic gift card of 25 US dollars after the interview.

\subsection{Qualitative Data Analysis} \label{sec:method:analysis}
We collected audio recordings from 21 interviews ranging from 40 to 92 minutes in length (average was 65 minutes).
We first leveraged Zoom's automatic transcription to generate initial transcripts and then manually reviewed all auto-generated transcripts for correctness according to the audio recordings.
Then, the first author structured all transcripts with multi-level section headings according to RQs to ensure effective navigation of transcripts in the coding process.
To rigorously assess the qualitative data, we combined both inductive and deductive coding approaches in our thematic analysis~\cite{Fereday2006,Clarke2015}. Four team members with prior qualitative data analysis experience were involved as coders to ensure internal reliability.
Our coding process included five steps: (1) The first author conducted the first round of inductive open coding and summarized emerging themes in the data;
(2) The research team discussed these themes with examples from the data and then create an initial codebook based on study RQs and themes identified in the first round coding.
(3) The research team used the codebook to perform the second round of coding, where the first author and three co-authors (secondary coders) double-coded all transcripts;
(4) The first authors discussed with three secondary coders individually to resolved all coding conflicts, which eliminated coder errors and further clarified a few codes. Inter-coder reliability measures are not necessary when coders reach full consensus~\cite{McDonald2019};
(5) The first author performed additional axial coding and selective coding~\cite{vollstedt2019introduction}
to synthesize high-level meta-themes.

\begin{figure}[H]
\centering
\begin{tikzpicture}[scale=0.070]
\draw[-] (-5,0) -- (105,0) ; 
\foreach \x in {0,15,30,45,55,70,85,100} 
\draw[densely dotted, shift={(\x,0)},color=black] (0pt,300pt) -- (0pt,0pt);
\foreach \x in {0,15,30,45,55,70,85,100} 
\draw[shift={(\x,0)},color=black] (0pt,0pt) -- (0pt,-0pt) node[below] 
{\scriptsize$\x\%$};
\draw (0,0) circle[radius=20pt];
\fill (0,0) circle[radius=20pt];
\draw (100,0) circle[radius=20pt];
\fill (100,0) circle[radius=20pt];
\fontfamily{ptm}{
\node[align=center,font=\scriptsize] at (103,5) {all};
\node[align=center,font=\scriptsize] at (-5,5) {none};
\node[align=center,font=\scriptsize] at (8,5.2) {a few};
\node[align=center,font=\scriptsize] at (22.5,4.9) {some};
\node[align=center,font=\scriptsize] at (37.5,4.5) {many};
\node[align=center,font=\scriptsize] at (50,5) {about\\half};
\node[align=center,font=\scriptsize] at (62.5,4.9) {majority};
\node[align=center,font=\scriptsize] at (78,5) {most};
\node[align=center,font=\scriptsize] at (92.5,5) {almost\\all};
}
\end{tikzpicture}
 \caption{Our terminology to describe theme frequencies}
    \label{fig:terminology}
\end{figure}
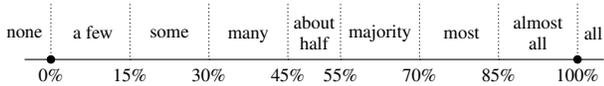

There is a growing recognition in the research community that the frequency of themes in qualitative research should not be interpreted quantitatively for generalization\cite{Emami2019,Habib2020}. Rather, the contribution lies in the identification of these themes and in-depth discussion of their implications.
We report frequencies of themes and codes when appropriate, but also adopt a consistent terminology used by Zhang et al.~\cite{Zhang2022} (Figure~\ref{fig:terminology}) to convey the relative prominence of themes. We archive our coded data including exact frequencies in Open Science Framework (see Appendix~\ref{sec:appendix:codeddata}).

\section{Participants and Baseline Questions}

\subsection{Sample and Demographics}
19 of 21 participants self-reported to be blind or legally blind, and two had low vision. 
Our sample is balanced across genders, age groups, and employment statuses (see Appendix~\ref{sec:appendix:demographics}). 
We over-sampled non-white participants (N=8) compared to the US demographic distribution, while two participants voluntarily reported as members of the LGBTQ+ community.
Our sample is biased towards people with high education levels as all participants had at least some college education.
Overall, our sample is relatively inclusive and mitigates common sampling bias with the BLV population.

\subsection{Participants' Technology Use} 
\label{sec:participants:tech}
Almost all participants reported having used computers (N=20) and phones or tablets (N=21).
Overall, we noticed an overall heavier reliance on mobile phones than computers. Specifically, two participants no longer use computers, citing the steep learning curve of computer screen readers and affordability considerations.
18 participants reported using various smart home devices, including smart speakers and other smart devices controlled through smart speakers.
Participants also mentioned using Victor Stream devices (N=4) and Braille Displays (N=5), both of which are assistive output devices with narrow functionality and thus excluded from our analysis due to their limited data privacy implications.

Participants used two major categories of digital technologies on their devices:
(1) \textbf{Assistive technology} that enables them to access various digital information, products, or services. All participants reported using screen readers on both computers (e.g., JAWS, NVDA) and mobile devices (e.g., VoiceOver, TalkBack). 18 participants used agent-based visual interpreter (VI) services, while 15 participants used AI-based visual-aid apps (e.g., Seeing AI). All participants also used voice assistants (e.g., Siri, Alexa) on their devices for accessibility.
(2) \textbf{General technology} that broadly refers to all devices, websites, apps, and digital products or services.

Additionally, our participants reported using computers mostly for formal activities such as work, banking, and information seeking. In contrast, they used their phones and/or tablets for broader online activities, including shopping, banking, communication, entertainment, and social media.
12 participants explicitly mentioned using websites or apps for their banking needs, but a few reported not using any online banking for security reasons.
Those who have smart speakers primarily use them for content, quick life help, and controlling other smart home devices.

\subsection{Data Privacy Awareness and Control} 
\label{sec:participants:awareness_control}
All participants identified at least one type of personal data being collected and knew advertising was a major purpose for data collection.
19 participants knew that digital technologies' data practices are pervasive. 
17 participants were aware of the existence of data privacy control options made available by digital technologies, many of whom reported having configured cookie settings, privacy settings, and app permissions.

However, nine participants expressed that the existing control options are ineffective.
Additionally, eight participants mentioned using some broadly defined privacy-enhancing technologies, including more private browsers or search engines and virtual private networks (VPNs). 
In summary, our participants are generally aware of data privacy and engage in some privacy control behaviors, which is similar to a recent sample with sighted Internet users in the US~\cite{Story2021}.

\subsection{Data Privacy Attitudes} 
\label{sec:participants:attitudes}
Our participants expressed multi-facet attitudes towards data privacy and we summarize the key themes below.

\noindent \textbf{Pervasive data practices and the hope for change.}
Participants described digital technologies' data practices as ``pervasive'' (N=9) and even ``intrusiveness'' (N=9).
Some participants believed such data practices should be limited or hoped for better privacy notice and choice mechanisms.

\noindent \textbf{Not concerned and resigned.}
Only two participants expressed explicit concerns about personal digital data privacy, while four were not concerned at all due to perceived low risks. Four participants expressed privacy resignation~\cite{Draper2017}, as P18 said: \textit{``It's very difficult to contain [data practices] no matter what message you try.''}

\noindent \textbf{Low data privacy expectations online and in daily life.}
Seven participants felt their online privacy was limited and 
two participants explicitly commented on their low privacy expectations as blind individuals, said P11:

\textit{``I think privacy is a very important thing...for blind people, we just don't get that privacy. There's nothing in my whole life that's private because being blind...somebody has to know everything to help me out with something.''}

\noindent \textbf{Neutral or positive attitudes towards data practices}. 
Seven participants were okay with reasonable use of their personal data such as ``providing or improving service'' but felt differently about targeted advertising.
Such a difference may derive from participants' opinions towards the data economy, the phenomenon that users' personal data is exchanged for access to a service and that data may subsequently be shared or sold to other entities.
For example, three participants have neutral views about the data economy, as explained by P17: \textit{``We are a capitalist society so companies need to make money.''}
Three participants mentioned personal benefits from targeted advertising, such as knowing about useful products.

Overall, the data reported in this section highlights the fact that BLV users must rely on assistive technology or sighted people (either from personal lives or interpreter services) to access online information, use other digital technologies, and manage their everyday lives, which invisibly impact their privacy perceptions and behaviors.
Our participants' privacy awareness and their diverse attitudes towards digital data privacy share many similarities with sighted users~\cite{kang2014privacy,acquisti2005privacy,Story2021}. 
\section{BLV Users' Risk Perception and Mitigation} \label{sec:results:RQ1}

\subsection{Perceived Security and Privacy Risks} \label{sec:results:RQ1:risks}
Our participants reported a broad range of security and privacy (S\&P) risks with digital technologies beyond the definition of data privacy presented to them (detailed in Appendix~\ref{sec:appendix:def}).

\noindent \textbf{General technology security risks.}  \label{sec:results:RQ1:risks:general}
15 participants mentioned the risks around personal financial data (e.g., social security numbers, credit card numbers, banking information).
Eight participants also mentioned data breaches that leak their sensitive data collected by legitimate companies, and some cited the Equifax data breach as an example.
Notably, the above-mentioned risks are security risks, not necessarily data privacy risks.
Our analysis showed that only seven participants clearly distinguished security risks from data privacy risks associated with data practices.
During the interview, we intentionally did not interrupt participants when they talked about security risks beyond data privacy.
Though not asked in our interview, six participants voluntarily shared their experience falling victim to financial crimes, from personal financial information being used by a neighbor to credit card numbers being stolen online.
Our participants' experiences revealed their vulnerability to financial crimes.
It is only natural for many participants to be more concerned about financial security than data privacy.

\noindent \textbf{Visual interpreter services: agents and companies.}  \label{sec:results:RQ1:risks:VI}
14 participants identified the agents working for visual interpreter (VI) services as a risk. P20 expressed her concern:
\textit{``When you call in, if they [agents] can't see it [the content to be interpreted] properly, they're supposed to ask permission before they snap a picture of it, but 99\% of agents will just snap a picture without asking you first now. It's your credit card sitting there; They've got your information.''}

Two participants pointed out the risks of how VI service companies store their personal data, as P24 explained: \textit{``There is the potential of my privacy being exposed because even though your information is stored in the cloud...those can also be compromised, and they may not know in time to inform you or to even make a solution for it.''}

Compared to VI services, participants were less concerned about privacy risks associated with AI-based visual-aid apps, as P17 said: \textit{``Seeing AI is not a problem because that's my phone reading it back to me. But Be My Eyes and Aira, I'm concerned about using them because it's a live person who may be reading sensitive information.''}

\noindent \textbf{Expanded physical-digital risks.}  \label{sec:results:RQ1:risks:physical}
Some participants mentioned physical-digital privacy risks, including shoulder surfing and eavesdropping by people nearby.
We found such physical-digital risks expanded to assistive technology and other scenarios. 
For example, P31 was concerned about using voice assistants in public: 
\textit{``One of the things I think is a real downside of the audio and the speech recognition style controls like Siri and Echo devices is that it forces you to speak information out loud...and that's a risk all by itself...if you forget other people [are] around sometimes.''}
Similarly, P11 mentioned a new category of physical-digital risk with VI services, she said  
\textit{``Not only [the agents], there are whoever happens to be in their houses...now they're all working from home, so you hear other people in the background, and I wonder, is this really secure?''}

\noindent \textbf{Risks around data practices.}
Participants mentioned risks associated with various data practices regarding both general technology (N=7) and assistive technology (N=2).
A few participants reported physical safety concerns related to location tracking, said P21: 
\textit{``If the data that's being collected could be shared with people that I haven't consented to, or could reveal things like my location and personal information about where I live, or might be expected to be in that create safety concerns.''}
Additionally, two raised concerns about unwanted surveillance, either by governments or companies. For example, P22 said,\textit{ ``data privacy issues now are becoming a political thing...if the wrong people figure out with your data you're not in the same political camp as they are''}, and pointed out the risk to civil liberty if being censored by digital technologies.

\noindent \textbf{More concerned about general technology.} \label{sec:results:RQ1:risks:moregeneral}
Mid-study, we added a follow-up question asking participants to compare their stated data privacy risks between general technology and assistive technology.
10 out of 13 participants who answered this question were more concerned about privacy risks with general technology. 
A few explained their rationale that assistive technology is a ``small market'' thus less likely to be targeted by attacks.
Also, some participants already mitigated assistive technology use (see section \ref{sec:results:RQ1:mitigation}) and thus were less concerned.
However, within assistive technology, VI services draw the highest level of concern, while screen readers are considered low-risk.

\noindent \textbf{Critical incidents around data privacy.} \label{sec:results:RQ1:risks:critical}
Our critical incident technique probed into the participants' recent experiences when they felt surprised, uncomfortable, or suspicious of certain data practices.
19 participants recalled at least one critical incident. 
Ten of them were surprised by unexpected features of certain apps or services, including Facebook's friend recommendations, auto-fill features on some websites, and Apple devices asking to share Wi-Fi passwords with contacts. 
Ten participants feel surprised or uncomfortable with targeted advertising, particularly by the speed and accuracy of these ads. Five of them specifically mentioned ads that appeared to be cross-platform, as P22 explained: 
\textit{``I’m on Amazon, and I searched for something, I would expect them to show me [the same] sort of things later on. I expect that. I don't expect Facebook to show it to me. And I've seen that happen.'' }
Additionally, many shared their experiences receiving unexpected spam/scam emails (N=5) or security incidents online (N=3).
Most of these reported critical incidents made participants realize how pervasively various digital technologies collect, use, and share their personal data.
However, many participants' responses reveal misconceptions or gaps in understanding about online behavioral advertising and cookies.

\subsection{Mitigation of Perceived Risks} \label{sec:results:RQ1:mitigation}
\noindent \textbf{Primary strategy: adjusted technology use} \label{sec:results:RQ1:mitigation:adjusted}
Adjusting their usage patterns of digital technologies is the most commonly mentioned risk mitigation strategy.
Four participants adjusted their general technology use, including limiting their social media usage to preserve online privacy (P22) and choosing mobile banking apps over web-based banking interfaces for better security (P24).
For assistive technology, ten participants mitigated the risks with visual interpreter services by intentionally not sharing sensitive information with Be My Eyes volunteers, as P13 said: \textit{
``With Be My Eyes...you do not use them to read credit cards to you that you got in the mail because your other one expired. Because they're not background checked, they're just people that really want to help...But with Aira, the agents are extensively trained; they're also background checked, and I think they're bonded or something, so I have had them read credit cards...''  
}

Such usage adjustments were not financially viable for all BLV users. A few participants in our study mentioned that cost is a key consideration when using agent-based VI services, as P14 explained:
\textit{`` I take advantage of it [Aira's promotion]. You don't have to pay for it for [participating] stores that you go shopping, you can get access [to Aira] and it doesn't cost in places like Walgreen [store].
But I also use Be My Eyes...That's a free one you don't have to pay for. They are volunteers from all over the world.''}

\noindent \textbf{Discontinued use and non-use.} \label{sec:results:RQ1:mitigation:nonuse}
Only five participants stopped using certain digital technologies or switching to alternatives out of privacy or security reasons.
These include deleting accounts on e-commerce sites due to security concerns (P11), stopping using social media apps out of privacy concerns (P12, P14), switching to new browsers because Internet Explorer is no longer maintained for security (P30), and adopting non-Google search engines to limit data exposure online (P27). 
However, no participants reported stopping using any assistive technologies for privacy and security reasons.
In contrast, seven participants chose the non-use strategy to avoid risks with digital technologies:
For general technology, a few did not use online or mobile banking to reduce financial risks. For assistive technology, a few mentioned that they avoided using agent-based VI services.

\noindent \textbf{Security and privacy practices.} \label{sec:results:RQ1:mitigation:practices}
Ten participants reported adopting good privacy and security practices for risk mitigation, such as using more private search engines or browsers, configuring privacy settings or mobile app permissions, using strong passwords, and being cautious wherever personal data is requested.
We also observed ~\textbf{varying levels of security and privacy knowledge}.
A few tech-savvy participants understood the benefits of using relatively effective security and privacy tools such as virtual private networks (VPNs). For example, P27 was confident about his security practices: \textit{``The security is so locked down around my credit cards that I think it takes an actual data breach to get them. I don't think it’s through the privacy settings.''}
In contrast, many participants \textbf{only followed generic S\&P advice}, and a few reported struggling with online privacy, including difficulty understanding how cookies work (P11).
A few participants also shared their frustration when security impeded accessibility, said P18: \textit{``I was doing [mobile] banking. I had someone put the app on [my phone], I could check my balance and inquiry transactions, but then they [the bank] changed it. every time you go on the site, you have to put a new password. The major problem is that I can't type it in. On the Apple phone, you can dictate everything except your password.''}

\noindent \textbf{Hypothetical dilemma with assistive technology.} \label{sec:results:RQ1:mitigation:dilemma}
Our hypothetical dilemma question forced participants to weigh up between accessibility and privacy: 17 participants reported that problematic data practices would affect their willingness to use certain assistive technology. 
However, the attitude may not stop them from using the assistive technology in such a dilemma.
Only eight firmly stated that they would stop using that assistive technology, most of whom felt confident in finding good alternatives.
Nine participants said it would depend on the risk level, as P19 explained: \textit{``No matter what you do,  we're still going to have a risk. It all depends on to what extent the information is shared and how it's being shared. ''} 
It also depended on how heavily they rely on the assistive technology, as P18 said: \textit{``It has to do with how much you need this device (technology). I'm totally blind, I don’t have any vision at all, and I live alone. So my need for the device is greater than someone else who maybe has some vision or lives with somebody.''}
Specifically, some participants mentioned giving up or switching screen readers would be the most difficult, and a few could not switch due to the limited available alternatives.
We notice that less technology-proficient participants were less likely to stop using or switch, said P11:
\textit{``It's always a concern, and if I found out that they did have a privacy (issue), they weren't really secure, or they were leaking information. it's hard to say, because if you are blind, you don't really have a whole lot of options. So do you take a chance and do it, or can you do without? Some things you just can't do without.''} 

Participants' responses revealed a high level of trust in assistive technology, with the exception of agent-based VI services.
For example, P25 said: 
\textit{``Especially with assistive technology, I operate with a very high level of trust. If they would do something that would erode my trust level, I would seriously consider changing things or making a complete switch to something different, if need be.''}
\section{BLV Users' Information (Seeking) Behavior} \label{sec:results:RQ2}

\subsection{Information Sources and Credibility}
\label{sec:results:RQ2:sources}
\noindent \textbf{Data privacy information sources.}
Participants reported various sources that they obtained data privacy information from.
Reputable news outlets (e.g., TV, radio, and print news) were the most reported source (N=13), followed by various online sources (N=11) and interpersonal sources (N=11) including tech-savvy friends and technical experts in the BLV communities.
A third reported obtaining data privacy information from privacy policies, terms of use, or user agreements. 
Six participants realized the existence of data practices based on their empirical experiences, such as receiving targeted and various spam/scam emails, to rationalize that their personal data was collected or shared by companies.

\noindent \textbf{Assessing information source credibility.}
\label{sec:results:RQ2:credibility}
Almost all participants reported certain criteria or preferences when assessing the credibility of information sources.
Our analysis revealed that \textbf{trust} played a critical role in their credibility assessment process.
15 participants believed \textbf{their trusted entities} could provide relatively credible information, including reputable news outlets and BLV organizations.
Nine participants placed trust in \textbf{the people} disseminating the information, including tech-savvy friends and influencers in the blind community.
A third of the participants reported sophisticated credibility assessment strategies, such as cross-referencing multiple sources and seeking primary sources.

\subsection{Privacy Policies as Information Source}
\label{sec:results:RQ2:PP}
\noindent \textbf{Experience with privacy policies.}
Eight participants mentioned privacy policies in their responses without prompts, and all participants said they heard about privacy policies before after the prompt. 
17 participants reported that they have read at least a few privacy policies before,
often when signing up for a new service or receiving privacy updates from companies.
Only two participants read privacy policies due to the perceived importance of the contents.
This result and \ref{sec:results:RQ2:sources} together indicate that privacy policies are an underutilized source by BLV users.

\noindent \textbf{Credibility and usability of privacy policies.}
\label{sec:results:RQ2:PPcredibility}
15 participants considered privacy policies a credible source for data privacy information because they are the official disclosure of companies' data practices. A few pointed out caveats with its credibility because privacy policies ``may change without notice'' and ``do not prevent security problems''. 
The remaining six participants thought privacy policies were not credible, because they ``lack accountability'' and credibility can ``vary by the companies''.
11 participants described difficulty reading privacy policies due to their length and vague languages, as P19 elaborated:
\textit{``It's a lengthy legal document so it's not like an exciting read to begin with...I feel that they're very vague...just very open to interpretation.''}

\subsection{Seeking and Non-seeking}
\label{sec:results:RQ2:seeking}
13 participants reported that they sought information regarding the data practices of digital technologies and the remaining eight participants did not, as detailed below.
\noindent \textbf{Seeking data privacy information}
Among 13 participants, eight actively looked for data privacy information multiple times, while five only mentioned one or two examples and admitted that rarely sought data privacy information unless they had a concern.
10 out of the 13 participants successfully found the data privacy information they were looking for, including clarification about data practices, confirmation of data practices mentioned elsewhere, available privacy controls, and detailed information of known data breaches. 
To our surprise, only three participants mentioned that their search outcome were less than satisfactory, citing challenges in finding relevant information (P11), exercising privacy controls (P12), and understanding privacy policies (P30).
Regarding information-seeking strategies, Google searches (N=6) and reputable news outlets (N=6) were the most commonly mentioned. Three participants reported using non-Google search engines for ``more neutral results''. A few participants also consulted privacy policies or terms of use (N=3), expert opinion (N=2), and trusted persons (N=2).
In summary, participants who sought information on data privacy generally succeeded in their search.

\noindent \textbf{Not seeking data privacy information.}
Among the eight participants who did not seek data privacy information, four reported that privacy was not a major concern to motivate their information seeking.
However, the lack of concern may derive from certain \textbf{misconceptions around data privacy}, as one marginalized participant said:
\textit{``(I didn't seek information) because it doesn't affect me. I don't have any information that is that important, that I would be upset about them collecting the information....I mean they can do anything with the little bit of money I got in the bank; I don't think it's important to them. I think I'm so glad that I’m down on the socioeconomic totem pole and that my information is not that important to them. But I think there are people whose information is.''}
This data suggests misconceptions may lead to an inaccurate assessment of data privacy risks and a failure to establish the appropriate level of concern.
Two participants thought it was not necessary. For example, P29, who has high technology literacy, felt that ``not much has changed in the data privacy landscape.''
Another participant admitted that ``fear'' prevented her from seeking such information, as she explained: \textit{``Information, mainly because it's a scary topic, like the more I know the less I want to be on the web.''}
Only one participant admitted that they did not seek such information due to the perceived difficulty because ``it's time-consuming and difficult'' (P31). 
In summary, \textbf{the lack of data privacy concern} reduced the necessity for participants to seek data privacy information, which is consistent with findings in ~\ref{sec:results:RQ1:risks} that data privacy risks were not participants' primary concern.

\section{Expectations for Privacy Q\&A Tools}\label{sec:results:RQ3} 
To mitigate participant response bias, we avoided priming participants with the idea of privacy Q\&A assistants by phrasing it as a hypothetical ``digital assitant'' (detailed in Section \ref{sec:method}).

\subsection{Expectations without Priming} \label{sec:results:RQ3:Free}
19 participants described at least one expectation for the privacy Q\&A assistant without priming, as detailed below.

\noindent \textbf{Good functionality.} Most participants expected good Q\&A functionality, which means the assistant should provide high-quality (N=9) and up-to-date (N=3) answers to their privacy questions in plain language (N6), as P15 described:
\textit{
``It would give answers to the questions we ask...in straight-out answers, it's not what we think it's [the data] going to be used for...No, it's gonna be, not assumptions, just cold hard facts.''}

\noindent \textbf{Accessibility by default.} 
Two participants explicitly expected the assistant to be accessible for BLV people. The other participants' follow-up comments confirmed that they implicitly assumed the assistant would be accessible by default.

\noindent \textbf{Advanced features.} Many participants elaborated on several advanced features, such as providing ``links to references'' in support of the answers(N=2), providing ``a general overview'' (N=2), and incorporating a mechanism to verify data privacy information.
Seven participants expected other advanced features, including attractive accent (assuming the assistant is voice-based), the capability to naturally interact with users, and personalized reminders of potentially risky data practices.

\noindent \textbf{Three prefer privacy experts over digital assistants.} P20 commented on the poor accessibility of other digital assistants: 
\textit{``I prefer it to be a human assistant because digital assistants can only give you answers that it's programmed to give. When you have something like Aira involved, it's person to person contact. Digital assistant is not enough to answer the concerns of a blind consumer.''}
Our analysis revealed that all three participants had negative user experiences with digital assistants. 
For example, P17 vividly shared a frustrating experience with us and commented:\textit{ ``I'm done dealing with digital assistants on several different websites.''}
This data indicates that prior negative experiences could impede BLV users' acceptance of digital assistants in the future. 

\subsection{Expectations for Privacy Q\&A Assistants}
\noindent \textbf{Cross-platform and multi-modality.} \label{sec:results:RQ3:devicemod}
While six participants preferred the privacy Q\&A assistant to be available on one device type (i.e, smartphones, computers), 15 participants hoped the privacy Q\&A assistant would be cross-platform and cross-device, as P13 commented: \textit{``probably all of them [the devices]...in whatever form or shape... but I think the assistant should be on all devices.''}

Regarding the interaction modality of the privacy Q\&A assistant, 15 participants want it to support both textual and auditory Q\&A experiences.
P26 explained his understanding of accessibility:
\textit{``I've always been interested in accessibility for the most people, so I would say both. By voice for people that are interested in something like that...but maybe a lot of people are deaf and unable to speak, so to have an alternative method like being able to type would definitely bring accessibility up a lot, and maybe even getting a response back in text for somebody that cannot hear.''}

Interesting, three participants strongly preferred the non-verbal interaction modality due to the limited accuracy of dictation on their devices, as P28 explained: 
\textit{``I’m more comfortable doing research by typing. I haven't had much success by voice searching on any device, I always prefer to type...[For answers, I prefer] text format that can be accessed by anyone, but with me, it would be the screen reader.''}
Notably, two participants pointed out \textbf{modality consistency} during their Q\&A interaction -- they would prefer to receive answers in the same modality as they asked questions.

\noindent \textbf{Preferred information sources.} \label{sec:results:RQ3:source}
Participants reported their preferred information sources from which the privacy Q\&A assistant should gather information.
13 participants preferred information from the first-party companies, about which the privacy questions were asked. 11 participants wanted information from the official legal documents including privacy policies and terms of use.
Many participants also mentioned reputable news outlets (N=6), other online sources (N=7), and organizations like privacy watch groups (N=6), while a few wanted the assistant to also look for actual regulations (N=2) as well as ratings of companies' privacy practices (N=2).
Notably, four participants would like the assistant to employ some verification mechanisms to ``constantly vet sources'' and ``compare what the company's saying to actually what has happened.''
This indicates participants valued both credible information sources and mechanisms (e.g.,\cite{Zimmeck2019MAPSSP,liu2021have} to verify the sources.

\noindent \textbf{Preferred developers.} \label{sec:results:RQ3:developers}
16 participants would trust \textbf{nonprofit organizations} to develop the assistant because they are ``neutral' and ``have guiding purposes''. Some cited Consumer Reports, the World Wide Web Consortium (W3C), and research universities as examples.
10 participants would trust \textbf{third-party companies} dedicated to developing the privacy Q\&A assistant. Participants believed they had the technical capability but ``no vested interest'' in the data practices being questioned.
Similarly, two participants preferred collaboration between two types of entities, as P31 said:\textit{ ``Either a nonprofit and a third party working together, one with the tech and one with the ethics, or it could be a governmental entity working with a nonprofit.''}
Interestingly, two participants strongly preferred first-party companies (e.g., big tech companies). P19 specified that she would trust Apple to develop the assistant because she valued Apple's product quality and customer support, even if the company may not be completely neutral.

Participants also discussed the entities they would not trust to develop the privacy Q\&A assistant. 
12 participants distrusted \textbf{first-party companies} that also engage in data practices being questioned, citing that the assistant may not be neutral because the companies have vested interest in the data practices.
\textbf{Governments} (N=10) were also unpopular due to participants' personal political opinions or their belief that governments cannot deliver good technology products.

\subsection{Benefits, Concerns, and Use Cases} \label{sec:results:RQ3:BCUS}
Assuming the privacy Q\&A assistant was developed by their preferred developers, participants elaborated on the benefits, concerns, and use cases for the privacy Q\&A assistant.

\noindent \textbf{Benefits.}
14 participants identified \textbf{easy access to trusted data privacy information} as a benefit. P13 explained:\textit{``I don't have to be scurrying, like a little kitty cat over the Internet, to try to find this information as much. The assistant can read the boring information and then give me a snapshot of what it has found.''}
Three participants mentioned that using the assistant could increase their awareness of data practices, said P24:\textit{ `` [Having] readily available, accurate information that was generated or produced by a group of people who have the knowledge...I, as a consumer, will be better informed about how my information is collected and shared. ''}
Some participants mentioned \textbf{benefits tied to their preferred developers}, including accessing neutral data privacy information (third-party companies or nonprofit organizations as developers) and enjoying highly-quality products (trusted first-party companies as developers).
A few participants mentioned that the assistant can increase their confidence when dealing with businesses or making privacy decisions online.
These results indicate that a privacy Q\&A assistant would benefit BLV users in many ways, regardless of whether they sought data privacy information or not.

\noindent \textbf{Concerns.} 
Even with their preferred developers, participants still expressed concerns about the privacy Q\&A assistant. 
Eight participants identified \textbf{the data practices of the assistant itself as a concern}, explained P24:
\textit{``The digital assistant is going to function like Siri or the smart speaker. Their constant availability accepting information via voice...because they're able to access more information and then that there might open up the channels for other people to have access to the information.''}
Note that our emphasis on data privacy in prior interview questions may have priming effect for this concern.
A few participants worried about the \textbf{long-term neutrality} of the privacy Q\&A assistant and the \textbf{quality of the answers} provided, said P14:\textit{``There would definitely have to be checks and balances so that the digital assistant wouldn't morph into something that could be used against the purpose of its existence.''} 
Only one participant explicitly voiced accessibility concerns from a blind user's perspective.

\noindent \textbf{Use cases.} 
14 participants reported that they would most likely ask questions about data privacy \textbf{before starting to use a new or unfamiliar digital technologies}, including \textit{``at the time of installing the app (P12)''} and \textit{``when connecting with a merchant that I've not used before (P14)''.}
Many participants said that they would like to use the privacy Q\&A assistant to check the digital technologies they currently use (N=4), obtain proof for companies' data practices (N=4), and when there are updates to companies' data practices (N=3).
Two participants said they would only use it \textbf{when there is a privacy concern}.
Notably, two participants expressed strong enthusiasm towards using the privacy Q\&A assistant, as P17 commented:\textit{
``[I would use it in] any place I have to put in something beyond my name. if I've got to put in my date of birth, my social security number, or any financial information linking any financial accounts to a certain company, I would definitely use that digital assistant to make sure that I know exactly what's happening with the data that I'm inputting.''}
\section{Discussion and Implications}

\subsection{Study Limitations}
We acknowledge several limitations of this study. First, a purely self-reported interview study has its inherent methodological limitations. To increase data validity, we employed techniques~\cite{Flanagan1954} to help participants recall their experiences and refined interview questions to evoke truthful answers.

Second, the qualitative interviewing method also restricted our sample. Though our sample size is on par or larger than prior security and privacy qualitative studies with BLV users, the findings may not generalize to all BLV users. 
Particularly, the US sample in this study cannot reflect the data privacy experience of BLV users worldwide, who may need to navigate more complicated consent procedures, such as the cookie banners mandated by the General Data Protection Regulation~\cite{GDPR}.
Nevertheless, compared to prior studies with BLV participants~\cite{Lau2015,Dosono2015,Ahmed2015,Pradhan2018,Hayes2019}, we recruited a more diverse sample across age, gender, and race/ethnicity, which yielded more inclusive perspectives and uncovered disparities within this underrepresented user group.

Third, due to the study focus, our interview questions for RQ3 cannot escape the participant response bias towards NLP-based privacy Q\&A assistants. This means our findings did not investigate other approaches that can also assist BLV users in navigating digital data privacy.
To mitigate this researcher-induced bias, we did not explicitly mention ``privacy Q\&A assistant'' or any NLP techniques in the interviews and carefully crafted our interview questions to focus on participants' needs and expectations for privacy question answering capabilities, as detailed in \ref{sec:method:questions}.

\subsection{Mitigating Risks for BLV Users}
\noindent \textbf{Heightened risks for BLV users.}
Besides confirming prior research that BLV users face high security and privacy (S\&P) risks when using digital technologies~\cite{Ahmed2015,Napoli2021,Akter2020}, this study revealed several \textbf{previously unstudied privacy risks}. Regarding VI services, we identify the privacy risks associated with VI companies' data practices and a new type of physical-digital risks caused by the VI agents' environment (\ref{sec:results:RQ1:risks:VI}). 
BLV users are vulnerable to S\&P risks because they de-prioritize S\&P concerns when  having to overcome accessibility challenges around digital technologies~\cite{Napoli2021}.
Our results further suggest participants' low levels of privacy concern (\ref{sec:participants:awareness_control}) and certain misconceptions (\ref{sec:results:RQ1:risks:critical}) may further expose them to heightened S\&P risks.

\noindent \textbf{BLV users' risk mitigation is insufficient.}
Our study is the first to articulate BLV users' risk mitigation strategies and behaviors.
Our participants primarily mitigate perceived risks by adjusting how they use digital technologies (\ref{sec:results:RQ1:mitigation}).
This is unique for assistive technology, including VI services, that some participants heavily relied on in their daily lives. With limited alternatives, BLV users \emph{had to weigh between the essential utilities and risk mitigation}.
We also found participants' risk perceptions affect their risk mitigation behaviors.
For example, most participants considered AI-based visual-aid apps low-risk (\ref{sec:results:RQ1:risks:VI}) because there seemed to be no human involvement. 
Such perception fails to consider S\&P risks associated with these apps' data practices and their underlying algorithms.
Our sample is skewed towards educated participants but their current risk mitigation strategies do not prevent them from many S\&P risks.

\noindent \textbf{Implications for mitigating S\&P risks.}
Our findings indicate the importance of providing accessible information to BLV users to enable accurate assessment of S\&P risks. This highlights the need for tools like privacy Q\&A assistants. 
Specifically, compared to sighted users, BLV users tend to mitigate risks through behaviors, presenting both opportunities and challenges for S\&P tools.
The opportunities lie in the potential impact of significantly improving their S\&P if they adopt effective tools.
However, our results also identify several challenges: S\&P tools should consider the unique risks faced by BLV users and their accessibility needs; these tools should be seamlessly integrated into their existing risk mitigation strategies without increased user burden.

\subsection{Creating Inclusive Privacy Tools}
This is the first study investigating BLV users' information behavior around data privacy information, which yielded novel findings on how they seek or do not seek such information (\ref{sec:results:RQ2:seeking}) and how they assess information sources (\ref{sec:results:RQ2:sources}).

\noindent \textbf{Generally successful information seeking outcomes}. Different from prior research~\cite{Mutula2016,Xie2021}, our participants who sought data privacy information were mostly successful and \textbf{reported little difficulty} during the process. Similarly, the main reason for not seeking such information was a lack of need or concern. However, it is impossible to know if those who claimed no need or concern fully understand the privacy risks or if they would feel the same way if they did.
Regarding data privacy information sources, besides news outlets and privacy policies, interpersonal and BLV-specific channels are particularly important for BLV users.

\noindent \textbf{Similar challenges but more patience with privacy policies}.
Our findings revealed how BLV participant felt and interacted with privacy policies (\ref{sec:results:RQ2:PP}), which is a major source for NLP-based privacy tools~\cite{Harkous2016,Kumar2020,Ravichander2019}.
We found that BLV participants \textbf{face similar challenges as sighted users} in comprehending privacy policies~\cite{Reidenberg2015,Vail2008,Oeldorf2019} but generally consider privacy policies a credible source. Surprisingly different from sighted users, many participants \textbf{had the patience} to read through multiple privacy policies.

\noindent \textbf{Implications for creating inclusive privacy tools}.
Our findings show that BLV users' information behaviors around data privacy exhibit both similarities and differences when compared to sighted users.
Particularly, we did not find prevailing evidence that BLV users face more challenges when seeking data privacy information beyond accessibility considerations. 
This cautions researchers about inaccurate assumptions that they might bring into their research or tool development.
Essentially, privacy tools should account for BLV users' information needs and align with their existing information behaviors.
For instance, a search tool may not benefit those who seldom seek data privacy information but NLP-based privacy tools capable of summarizing key information from privacy policies~\cite{Kumar2020,zaeem2020privacycheck,keymanesh2020toward} can alleviate the usability frustrations for both sighted and BLV users.
This calls for more research to understand underrepresented user groups' privacy perceptions and behaviors to design more inclusive privacy tools.

\subsection{Strengthening Trust with BLV Users}
Trust is a meta-theme that impacts BLV participants' technology selection and their approach to data privacy information.

\noindent \textbf{Trust in entities and people.}
Participants' answers to the only trust-focused interview question (\ref{sec:results:RQ3:developers}) suggest that BLV users' trust in developers may impact their decision to adopt S\&P tools.
Our results showed that BLV users \textbf{distrust} big technology companies and governments to create transparent privacy tools.
In contrast, they tend to trust companies that provide assistive technology and products with outstanding accessibility features (\ref{sec:results:RQ1:mitigation:dilemma} \& \ref{sec:results:RQ3:developers}).
They also trust organizations and experts of the BLV community as well reliable friends and family members, because they believe these organizations and people have their best interests in mind.

\noindent \textbf{Trust in risk and credibility assessment.}
BLV participants considered screen readers and AI-based visual-aid apps low-risk because these digital technologies do not directly expose data to strangers (human agents). Notably, our participants seemed to trust AI-based tools and did not express concerns about relying on these AI-based tools, potentially leading to overlooked AI-induced S\&P risks. 
Additionally, in line with \cite{Hayes2019}, we find that BLV participants' existing trust also impacts how they assess the credibility of data privacy information. The participants who did not use sophisticated credibility assessment criteria tended to believe information mostly from the people or channels they already trust(\ref{sec:results:RQ2:credibility}). 

\noindent \textbf{Implications for building trust with BLV users.}
Trust is integral to BLV users' interactions with digital technologies, assessment of information credibility, and evaluation of S\&P risks. 
S\&P tools should establish trust with BLV users, where ensuring S\&P tools' accessibility is an essential starting point.
Collaborating with BLV organizations can further strengthen trust with BLV users.
Moreover, under the backdrop that popular generative AI applications like ChatGPT often provide false answers~\cite{Stokel2023}, it is particularly crucial for tools like privacy Q\&A assistants to build and strengthen trust with BLV and other marginalized users.

\subsection{Expectations for Privacy Q\&A Assistants}
This is the first study exploring BLV participants' expectations for potential privacy Q\&A assistants, enriching the growing research to design user-centered privacy assistants~\cite{Stover2023,Seymour2020,Colnago2020} with a focus on accessibility and the need of BLV users.
While our participants may prioritize data security over data privacy(\ref{sec:results:RQ1:risks:general}), they clearly recognized the potential benefits of privacy Q\&A assistants and expressed interest in them(\ref{sec:results:RQ3:BCUS}).

\noindent \textbf{Rethinking accessibility.}
Our BLV participants expected the assistant to be ~\textbf{accessible by default}, meaning they could access the assistant across devices and platforms and via different modalities (\ref{sec:results:RQ3:devicemod}).
Different from prior research showing that BLV users found voice-based apps particularly useful~\cite{Pradhan2018}, our participants preferred interacting with the assistant in \textbf{multiple modalities}. Some also favored \textbf{textual input and output} so that they can ask more precise questions and obtain more comprehensive answers. This highlights the importance to extend existing NLP Q\&A evaluations~\cite{Ravichander2021a,HakkaniTur2022,Faisal2021} to account for inputs from different modalities like speech.

\noindent \textbf{The importance of quality and trustworthiness.}
Most participants implicitly assumed that the privacy Q\&A assistant could provide high-quality answers to their privacy questions (\ref{sec:results:RQ3:Free}).
Some also imagined advanced features to enhance answer credibility, including the ability to cross-reference sources and to verify data practices objectively.
In addition, many participants generally emphasized the importance of having trustworthy assistants with their best interest in mind when providing answers.
Given that three participants still prefer human experts due to prior negative experience with online chatbots, the assistants would need to provide high-quality and trustworthy answers to swing their perspectives.

\noindent \textbf{Implications for privacy Q\&A assistants.}
These findings underscore the significance of default accessibility for privacy Q\&A assistants and broader S\&P tools. To achieve optimal accessibility, developers should determine the best modality combination based on different user groups' accessibility needs~\cite{Himmelsbach2015}, which typically supports multiple modalities.
Moreover, BLV users' expectations for functionality drives their interest in privacy Q\&A assistants. To meet such expectations, potential areas of research include applying NLP techniques to consolidate relevant information from both privacy policies and other credible sources (e.g., privacy regulations and news), and advancing verification mechanisms to validate the data practices disclosed in privacy policies.

\subsection{Towards Equitable Privacy}
\noindent \textbf{Digital divide.}
The digital divide generally refers to unequal access to digital technology among populations. 
This was seen in several unemployed or racially minoritized participants.
For example, one participant never learned to use computers due to the lack of affordable screen reader training (\ref{sec:participants:tech}) and another had to use the free but relatively risky VI services Be My Eyes due to financial considerations (\ref{sec:results:RQ1:mitigation}).
We also observed that marginalized BLV participants in our sample heavily relied on mobile phones to access digital technologies.
This directly increases the S\&P risks faced by marginalized BLV users and limits their risk mitigation options.

\noindent \textbf{Technology literacy.}
Our results reflected the inequality caused by disparities in technology literacy.
A few participants enjoyed the security provided by paid privacy-enhancing technology like VPNs (\ref{sec:participants:tech}), while one marginalized participant struggled with entering passwords on their iPhone that prevented them from using mobile banking(\ref{sec:results:RQ1:mitigation:practices}).

\noindent \textbf{Implications for S\&P research.}
Our inclusive sample yielded perspectives on equity and inclusion even among BLV users.
While technology alone cannot resolve the disparity in security and privacy resulting from societal factors, it is essential for the S\&P community to acknowledge the presence of such inequality.
We should actively work towards developing accessible, inclusive, and equitable S\&P tools that can benefit the widest possible spectrum of users.

\section{Conclusion}
We presented an in-depth interview study aimed at understanding BLV users' data privacy risk perception and mitigation strategies, their information behaviors related to data privacy, and their expectations for privacy Q\&A tools that could assist them in navigating data privacy information.
This study yielded rich findings and implications around usability, accessibility, trust, equity, and S\&P risk mitigation for BLV users and beyond.
We want to conclude this paper with one participant's quote that uncovers the core value of accessibility:
\begin{quote}
\textit{``I know the survey [interview] is supposed to be for the blind...but really what you're talking about to me is a survey [interview] everybody should take...because I think what you're talking about [the privacy Q\&A assistant]...will benefit everybody, and you know a lot of the cases in the community of people with disabilities, we try to find things that are going to benefit the whole populace.''}
\end{quote}

\section*{Acknowledgments}
We want to thank our blind consultants Dr. Cynthia Bennett and Chancey Fleet for their invaluable advice and the National Federation of the Blind for their help with participant recruitment. 
This study is funded by the US National Science Foundation under the project ``Automatically Answering People's Privacy Questions'' (CNS-1914444 \& CNS-1914486). 
The US government is authorized to reproduce and distribute reprints for government purposes not withstanding any copyright notices thereon. The views and conclusions contained herein are those of the authors and should not be interpreted as representing official policies or endorsements, either expressed or implied by NSF or the US government. 


\bibliographystyle{plain}
\bibliography{BLV.bib}

\appendix
\appendix
\section{Interview Questions} 
\label{sec:appendix:questions}
We list all key interview questions and describe some (but not all) follow-up probing questions due to the page limit.

\paragraph{Baseline questions:}
(1) What electronic devices do you typically use to access digital information?
(2) What assistive tools or technologies and other websites/apps/services do you use on this device? (for each type of device mentioned)
(3) Do you know what types of personal data is collected when you use these digital technologies on your devices?
(4) Considering the extend of data practices discussed just now, what are your general thoughts about data privacy in regarding digital technologies?

\paragraph{RQ1: Risks perception and mitigation:} 
(1)  Please think about the digital tools/technologies that you use, can you think of some of the potential risks around your personal data privacy?  (Follow-up questions about comparing risks between general technology and assistive technology).
(2) \textit{[Critical incident]} Please think about the past month when you used digital tools/technologies, were there any situations that you felt surprised, uncomfortable, or suspicious about how these tools/technologies use your personal data?
(3) Have you ever stopped using a tool/technology or switched to an alternative tool/technology out of data privacy concerns?
(4) \textit{(Hypothetical dilemma question)} If you find out an assistive technology that you use handles your data in a way that you feel uncomfortable with, would it affect your willingness to use the technology?

\paragraph{RQ2: Information (seeking) behaviors:}
(1) How did you typically come across or get information about data privacy in the digital world? (follow-up questions: where, how, information sources , and perceived credibility of information sources)
(2) Have you ever tried to find any information about data privacy? (If yes, follow up with where, how, and information sources; If not, ask why)
(3) \textit{(When participants did not mention privacy policies as a source)} Are you familiar with privacy policies? (follow up on perceived credibility for privacy policies)

\paragraph{RQ3: Expectations for Q\&A tools:} 
(1) You mentioned that you used [X]. Please imagine if there is an expert who can answer any questions around data privacy for [X]. What kinds of questions would you ask this expert about [X]? ([X] is a digital tool/technology mentioned by participants; up to 4 tools/technologies of different categories are asked here) --- \textit{Note: the data from this set of questions is not reported due to the scope and focus of this paper.}
(2) Please imagine if the expert we discussed above is a digital assistant that can provide you with information around data privacy, how would you like the privacy assistant to be? (First, we let participants freely describe their imagined digital assistant without priming; then we asked follow up questions on preferred devices, modality, sources, developers, benefits, concerns, and use cases.)

\section{Definitions Given to Participants} \label{sec:appendix:def}
\noindent \textbf{Digital technologies}: During this interview, I will use the general term ``digital technologies'' to refer to all the websites, apps, web services, and assistive tools/technologies that you mentioned. Does it sound okay to you?

\noindent \textbf{Data practices}: As you may know, when you use these digital tools/technologies on your devices, your personal data is often collected, used, or shared by these tools/technologies.

\noindent \textbf{Data privacy}: Data privacy concerns the handling of personal data by different entities, such as if the handling is appropriate and if it’s in compliance with laws. The handling of personal data includes a variety of practices, such as how data is collected, used, or stored, whether data collectors share or sell the data to others. 
The focus of this interview study is data privacy with digital technologies, or digital data privacy.

\section{Coded Data} \label{sec:appendix:codeddata}
We archive the coded data in this Open Science Framework repository: \url{https://doi.org/10.17605/OSF.IO/K9FV6}

\section{Participants' Demographic Distribution} 
\label{sec:appendix:demographics}
\begin{table}[ht]
\centering
\small
\setlength{\aboverulesep}{0pt}
\setlength{\belowrulesep}{0pt}
\begin{tabular}{lrlrlrlr}
\toprule
\multicolumn{2}{c}{\cellcolor{gray!20} Gender} &
\multicolumn{2}{c}{\cellcolor{gray!20} Education} \\ \midrule
Female  & 52.4\%  & Some college    & 38.1\% \\
Male    & 47.6\%  & College degree  & 33.3\% \\
        &         & Graduate degree & 28.6\% \\
\midrule
\multicolumn{2}{c}{\cellcolor{gray!20}Race/Ethnicity} &
\multicolumn{2}{c}{\cellcolor{gray!20}Age Group} \\
\midrule
White     & 61.9\%  & 18--29   &  9.5\% \\
Black     & 19.0\%  & 30--39   & 19.0\% \\
Asian     &  9.5\%  & 40--49   & 19.0\% \\  
Hispanic  &  4.8\%  & 50--59   & 28.6\% \\
Mixed     &  4.8\%  & 60--69   & 14.3\% \\ 
          &         & 70+      &  9.5\% \\ 
\midrule
\multicolumn{4}{c}{\cellcolor{gray!20}Employment Status} \\
\midrule
\multicolumn{2}{l}{Fully employed}      & \multicolumn{2}{r}{23.8\%} \\
\multicolumn{2}{l}{Partially employed}  & \multicolumn{2}{r}{14.3\%} \\
\multicolumn{2}{l}{Self-employed}       &  \multicolumn{2}{r}{9.5\%} \\
\multicolumn{2}{l}{Unemployed}          & \multicolumn{2}{r}{28.6\%} \\
\multicolumn{2}{l}{Retired}             & \multicolumn{2}{r}{19.0\%} \\
\multicolumn{2}{l}{Homemaker}           &  \multicolumn{2}{r}{4.8\%} \\
\bottomrule
\end{tabular}%
\label{tab:demographics}
\end{table}


\end{document}